\documentclass[a4paper,11pt]{article}
\pdfoutput=1
\usepackage{jcappub}
\usepackage{graphicx}
\graphicspath{{figures/}}
 \usepackage{caption}
\usepackage{mathrsfs}
\usepackage{amssymb}
\usepackage{textcomp}
\usepackage{amsmath, float}
 \usepackage{times}
\usepackage{color}
\usepackage{enumitem}

\title{\boldmath Constraining cosmic curvature by using age of galaxies and gravitational lenses}

\author[a,1]{Akshay Rana,\note{Corresponding author.}}
\author[b]{Deepak Jain,}
\author[a]{Shobhit Mahajan,}
\author[a]{Amitabha Mukherjee}

\affiliation[a]{Department of Physics and Astrophysics, University of Delhi, Delhi 110007, India}
\affiliation[b]{Deen Dayal Upadhyaya College, University of Delhi,
Sector-3, Dwarka, Delhi 110078, India}

\emailAdd{arana@physics.du.ac.in}
\emailAdd{djain@ddu.du.ac.in}
\emailAdd{shobhit.mahajan@gmail.com}
\emailAdd{amimukh@gmail.com}

\abstract{ We use two model-independent methods to constrain the curvature of the universe. In the first method, we study the evolution of  the curvature parameter ($\Omega_k^0$) with redshift by using the observations of the Hubble parameter and transverse  comoving distances obtained from the age of galaxies. Secondly, we also use  an indirect method based on the mean image separation statistics of gravitationally lensed quasars. The basis of this methodology is that  the  average image separation of lensed images will show a positive, negative or zero correlation with the  source redshift  in a closed, open or flat universe respectively. In order  to smoothen the datasets used in both the methods, we use a non-parametric method namely, Gaussian Process (GP). Finally from first method we obtain $\Omega_k^0= 0.025\pm0.57$ for a presumed flat universe while the cosmic curvature  remains constant throughout the redshift region  $0<z<1.37$ which indicates that the universe may be homogeneous. Moreover, the combined result from both the methods suggests that the universe is marginally closed. However, a flat universe can be incorporated at $3\sigma$ level.}

\begin{document}
\maketitle
\section{Introduction}
\label{sec:intro}

Cosmic curvature  is a fundamental parameter in cosmology. It plays a  crucial role in the  evolution of universe. Furthermore, the combined observational constraints by SNe Ia, BAO, Hubble Data and WMAP survey also suggest   that $\Omega_k^0$ is indeed very small \cite{komatsu}.  Recently,   Planck survey  has given  very precise bounds on curvature density,  $\Omega_k^0= 0.000\pm 0.005$ ($95\%$, Planck TT+lowP+lensing+BAO) \cite{planck}. Interestingly, in order to find the solution of cosmological constant and coincidence problem, Shaw \& Barrow \cite{barrow} also predicted the spatial curvature parameter $\Omega_k^0= -0.0055$ which is also consistent with Planck result. Though the flatness of universe obtained from the cosmological observations is almost at an unprecedented level of precision, its interpretation still depends on the  fact that the background cosmology is described by homogeneous and isotropic  FLRW metric.  In order to test the FLRW metric and curvature of the universe, Clarkson et. al. proposed that any detection of redshift dependence of curvature parameter would imply presence of non FLRW cosmology \cite{clarkson2007}.\\

Another  important issue  is the presence of the  geometric degeneracy between dark energy equation of state parameter $w(z)$ and cosmic curvature $\Omega_k^0$. This degeneracy prevents us from understanding the nature of dark energy. To circumvent  this problem, most of the work  neglects  $\Omega_k^0$ (by advocating that several observations put very small bounds on $\Omega_k^0$) and attempts  to find bounds on $w(z)$ independently. However, it has been found that ignoring $\Omega_k^0$ leads to errors in the reconstruction of $w(z)$ and even a small value of $\Omega_k^0$ induces  a large effect at higher redshifts $z \simeq 1$ and the inclusion of cosmic curvature in the analysis offers a  wide range of dark energy models which can explain the accelerating universe. \cite{pavlov13,farooq13,chen16,farooqRatra16}. Hence it seems inconsistent to neglect the contribution of $\Omega_k^0$, howsoever small it might be. \cite{chris}. Moreover many dark energy tests  such as the Integrated Sachs-Wolfe (ISW) effect and cluster surveys  are also very sensitive to the assumption about  $\Omega_k^0$ \cite{isw}. In addition, weak lensing and baryon accoustic oscillation (BAO) datasets have also been used to constrain the cosmic curvature based on the distance sum rule \cite{berstein2006, rasanan2015}. However,  the efficiency of this test is limited because of the large uncertainties  in the weak gravitational lensing datasets. Many alternate theories (theories based on the anthropic principle \cite{anth}, the Inhomogeneous universe \cite{thomas}  and the Irrotational dust universe \cite{harald} etc.)  have been proposed to address the fine tuning of parameters and the presence of dark energy. Furthermore, a precise measure of cosmic curvature can act as an excellent test for several inflationary models of the universe supporting flat \cite{inf1,inf2,inf3} or open universe \cite{gott82,ratra94,ratra95,kamio94,bucher94,lyth95,yama95,gorski95}.Hence it is very important to constrain  $\Omega_k^0$  in a model  independent  manner.

In  this paper, we use  two different  model independent  techniques to estimate the curvature of the universe. In Method I,  we  determine $\Omega_k^0$ by using the  independent dataset of Hubble parameter $H(z)$ and the transverse comoving distance obtained by using the age of  galaxies \cite{holanda2016}. A prominent feature of this analysis is that  it doesn't  rely on any assumption of the underlying cosmological model. If we find  any deviation in the evolution of cosmic curvature then it is an indication of  new physics . We apply Gaussian Process (GP) smoothing technique to the data points to infer any violation of FLRW based cosmological models. In Method II, $\Omega_k^0$ is constrained by using the statistical properties of gravitational lens systems. Gott, Park and Lee  showed that for the Singular Isothermal Spherical (SIS) model of  galaxies, the mean image separation $<\Delta \theta>$ is independent of the source redshift if the universe is assumed to be flat in FLRW universe \cite{gott89}.  We use an  updated dataset of $44$ gravitational lenses from the  DR7 SDSS gravitational lens dataset \cite{Inada2012}. We  again apply the  non-parametric  technique of Gaussian Process on  the image separation data points to constrain the cosmic curvature.\\

The structure of the paper is as follows: In Section \ref{Method I} , we discuss  Method I based on the determination of  the cosmic curvature in a model independent manner  and  methodology to obtain the transverse comoving distance from the age of galaxies.  In Section \ref{Method II}, we  explain  Method II  which is based on the image separation statistics of gravitationally lensed quasars. We  analyze the results  in Section \ref{Results}.

\section{Method I : Null test of cosmic curvature using $H(z)$  and age of galaxies} 
\label{Method I}
This method provides a model independent test of spatial  curvature of the universe.  It  is based on a geometrical relation between the  Hubble parameter $H(z)$, transverse comoving distance $r(z)$ and its first derivative $r'(z)$ . It enable us to  check  whether the present  curvature density  is independent of the redshift  of measurement or not.This test was firstly proposed by  Clarkson et.al. \cite{clarkson2007} and further used by  several authors \cite{sapone, mortsell2011,cai2016,li2014,wei2016,wang2016,recent2016} where independent observations of Hubble parameter and transverse comoving distances are used to constrain the  cosmic curvature.  The relation is given as ;

  \begin{equation}
 \Omega_k^0= \frac{H(z)^2 r'(z)^2-c^2}{H_0^2 r(z)^2}
\label{ok}
 \end{equation}

It  measure  the present curvature density by using the independent observations of  $H(z)$ and transverse comoving distance at one single redshift. In any FLRW universe, the curvature parameter remains independent of redshift of measurement.  In other words,  If the  present curvature parameter is found to be dependent on the redshift of measurement,  then there is a need to explore the models beyond FLRW metric. Eq.\ref{ok} provide us  the simplest  way to analyze the present cosmic curvature by using the independent measurements of $H(z)$, $r(z)$ and $r'(z)$ at a given redshift.

\subsection{Estimating the transverse comoving distance using the age-redshift dataset of galaxies}\label{ss1s1}

To find the transverse comoving distance $r(z)$ and its derivative $r'(z)$, we use $32$ data points of age of galaxy ($0.11<z<1.84$) plus one data point from the Planck survey for the  present age of the universe. Out of these $32$ data-points, $20$  are  red galaxies from the Gemini Deep Survey (GDDS)\cite{gdss}, $10$ are early field type galaxies \cite{treu} whose ages are obtained by using SPEED models \cite{jimenez} and there are  two radio galaxies LBDS 53W069 and LBDS 53W091 \cite{simon}.\\

The transverse comoving distance $r(z)$ can be obtained from the age-redshift data of galaxies \cite{holanda2016}.   FLRW cosmology enables us to write down  the transverse comoving distance $r(z)$  in the form of the first derivative of the age of the galaxy,

\begin{equation}
r(z) = \begin{cases}
 \dfrac{c}{H_0\sqrt{|\Omega_{k}^0}|}\sinh{[H_0\sqrt{|\Omega_{k}^0}|}\int\limits_z^0 (1+\acute{z})\frac{dt}{d\acute{z}} d\acute{z}] & \mbox{for $\Omega_{k}^0>$0 }\\
{c}\int\limits_z^0 (1+\acute{z})\frac{dt}{d\acute{z}} d\acute{z} & \mbox{for $\Omega_{k}^0=$0}\\
 \dfrac{c}{H_0\sqrt{|\Omega_{k}^0}|}\sin[{H_0\sqrt{|\Omega_{k}^0}|}\int\limits_z^0 (1+\acute{z})\frac{dt}{d\acute{z}} d\acute{z}] & \mbox{for $\Omega_{k}^0<$0 }\\
 \end{cases} \label{rdef}
\end{equation}

where $t$ is the age of universe at redshift $z$. If one can determine   $\frac{dt}{dz}$  at the required redshift then one  can easily reconstruct the value of transverse comoving distance $r(z)$. For this purpose we use an  age-redshift dataset of $32$ old, passive galaxies distributed over the redshift interval of $0.11<z<1.84$. Further  we add an incubation time $t_{inc}=1.50 \pm 0.45 \text{ Gyr}$ that accounts for the time from the  beginning of the universe to the formation of the first galaxy. This is the mean value of the incubation bound obtained by Wei et. al. \cite{wei} using the  age of a galaxy sample in a flat $\Lambda$ CDM model. However, a  $12\%$ error is given  in the analysis of  measurement of age of galaxies which we take as the uncertainty in measurement \cite{dantus1,dantus2,sumushia}.\\

\begin{figure}[ht]
\centering
\includegraphics[width=10cm, height=6cm]{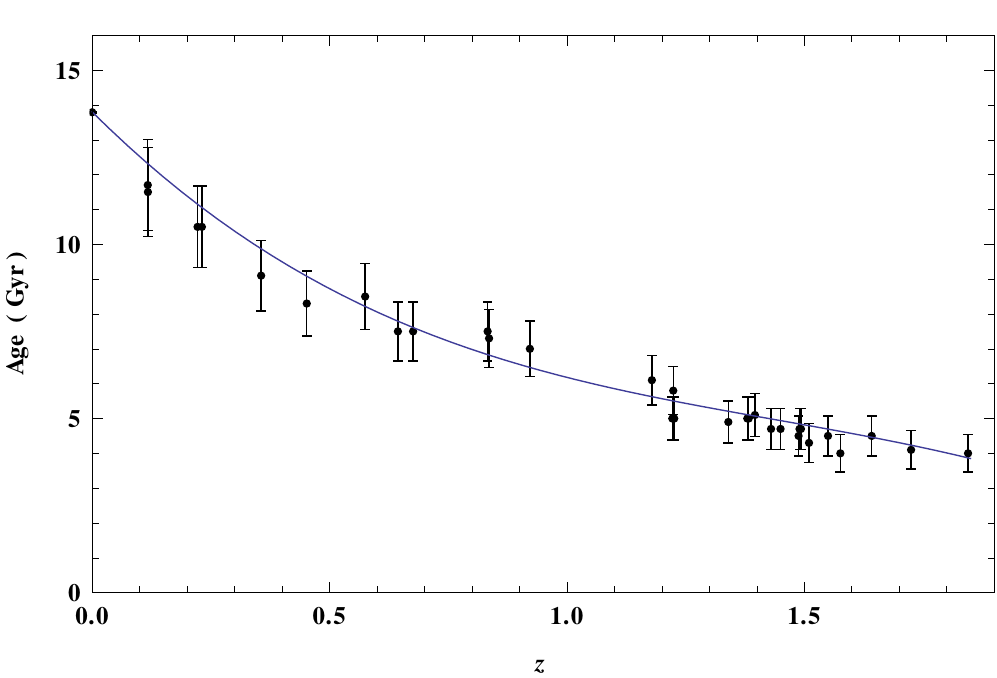}
\caption{\label{fig4} \small Age of galaxy $t(z)$ vs $z$ . The black points are the original data of $32$ points added with an incubation time $t_{inc}= 1.50\pm 0.45 \text{ Gyr}$. The blue line is obtained from the fit by using a third degree polynomial.}
\end{figure}

The  present age of universe is taken to be $13.790\pm0.021$ Gyr  obtained by Planck survey with a joint analysis of CMB + BAO + SNe Ia + $H_0$ \cite{planck}. Once we have the dataset, we fit this data using a third degree polynomial $t(z)= A+Bz+Cz^2$. The best fit values turn out to be (in Gyr): $A= 13.78\pm0.01$, $B=-10.65\pm0.84$, $C= 2.98\pm0.34$ with a  $\chi^2_{\nu}$= $0.41$. The best fit curve is shown in  Fig. \ref{fig4}.\\

On differentiating the polynomial, we obtain $\frac{dt}{dz}= B+2Cz$. Now here we choose $\Omega_k^0=0$ and by using Eq. \ref{rdef}, we obtained the transverse comoving distance;

\begin{equation}
 r(z) = c\left[   -B\left(z+\frac{z^2}{2}\right)-C\left(z^2+\frac{2z^3}{3}\right)-D\left(z^3+\frac{3z^4}{4}\right)\right]
   \label{rz}
   \end{equation}

and on further differentiating it, we get
\begin{equation}
     r'(z)= c\left[   -B\left(1+z\right)-2C\left(z+z^2\right)-3D\left(z^2+z^3\right)\right]
    \label{rpz}
   \end{equation}

It is important to note  that transverse comoving distance $r(z)$ also depends on the value of $\Omega_k^0$ so it can be derived using different values of $\Omega_k^0$. However, our aim here is only to study the evolution of $\Omega_k^0$ with redshift $z$. It helps  to check the consistency of the  FLRW metric as well as  homogeneity of the  universe. Therefore, any choice of $\Omega_k^0$ to derive the transverse  comoving  distance would fulfill our purpose.

In order to check the variation of $\Omega_k^0$ by using the observations at different redshift, we further need the Hubble data along with transverse comoving distance $r(z)$ and its first derivative $r'(z)$.

\subsection{Hubble dataset}\label{s2}

We use a recent dataset of $H(z)$ consisting of 38 data points which are derived using various observational methods and datasets [see Table \ref{tab}]. It is important to stress that in order to derive the value of $\Omega_k^0$, both the datasets ( i.e. $H(z)$ and age of galaxies used to calculate transverse comoving distances)  must be independent of each other. Therefore we use the following methodology.

$\bullet$  It is observed that in 	the Hubble data set,  $9$ data-points are obtained  by using the same $32$ age of galaxy data points or their subsamples, which  are used to derive the transverse comoving distances in our work \cite{simon}. So we remove these nine  points from the Hubble dataset. These points are indicated by $R^a$ in  Table \ref{tab}.

$\bullet$ Further in the Hubble dataset, all $38$ measurements of $H(z)$ are not completely independent. Three measurements taken from Blake et.al (2012) \cite{hub8} (See S.No. 16, 22 ,25 in Table \ref{tab}) are correlated with each other  and three measurements taken from Alam et.al (2016) \cite{alam} (See  S.No. 11, 20, 23 in Table \ref{tab}) are also correlated. All these points are derived from the BAO measurement. However, Farooq et al (2016) \cite{farooqRatra16} have recently observed that the   inclusion of the off-diagonal elements of the correlation matrix of these $H(z)$  measurements will produce a small effect on the likelihood function used for the parameter estimation. Following the same reasoning, we therefore  treat these observations as independent by neglecting the off diagonal terms of correlation matrix in our analysis.

$\bullet$  The upper limit of redshift in the age dataset is $z<1.84$. Hence we restrict our analysis till the redshift $z= 1.84$ and we drop the three  $H(z)$ observations at $z=1.965$ from Moresco et. al (2015) \cite{moresco15}, $z=2.34$ from Delubac et.al (2015) \cite{delubac} and $z=2.36$ from  Font-Ribera et.al. (2014)\cite{font} from the H(z) dataset.

Finallly, we are left with $26$ observations of $H(z)$. By using these two independent datasets,  the Hubble parameter and the transverse comoving distances derived  by using the age of galaxies, we  reconstruct a dataset of the  cosmic curvature $\Omega_k^0$. The corresponding error bars are derived by using the standard error propagation method.

\begin{equation}
\sigma_{\Omega_k^0}^2= 4(\Omega_k^0)^2 \left[ \left( \frac{\sigma_r}{r}\right)^2 +\left( \frac{\sigma_{H_0}}{H_0}\right)^2  \right] +4\left[(\Omega_k^0)^2+ \frac{c^2}{(H_0 r)^2}\right]\left[ \left( \frac{\sigma_H}{H}\right)^2 +\left( \frac{\sigma_{r'}}{r'}\right)^2  \right]
\label{sigma1}
\end{equation}

where $\sigma_r$, $\sigma_{r'}$,  $\sigma_H$ and  $\sigma_{H_0}$ are the errors in the transverse comoving distance $r(z)$, its first derivative $r'(z)$, the Hubble parameter $H(z)$ and the present value of Hubble constant $H_0$ respectively.

The value of $H_0$ to be used in the analysis is another issue. Chen \& Ratra 2011) \cite{chenH011} have obtained a value $H_0= 68 \pm 2.8$ Km/sec/Mpc which is widely supported by global observations \cite{sievers13,hinshaw13,aubourg14,planckh0,ychan16}. On the other hand, using SNe Ia data, Riess et. al \cite{riess} have found  $H_0= 73.24 \pm 1.74$ Km/sec/Mpc. We carry out our analysis using both these values of the Hubble constant.

\begin{figure}[ht]
\includegraphics[height=5.5cm,width=7.5cm,scale=4]{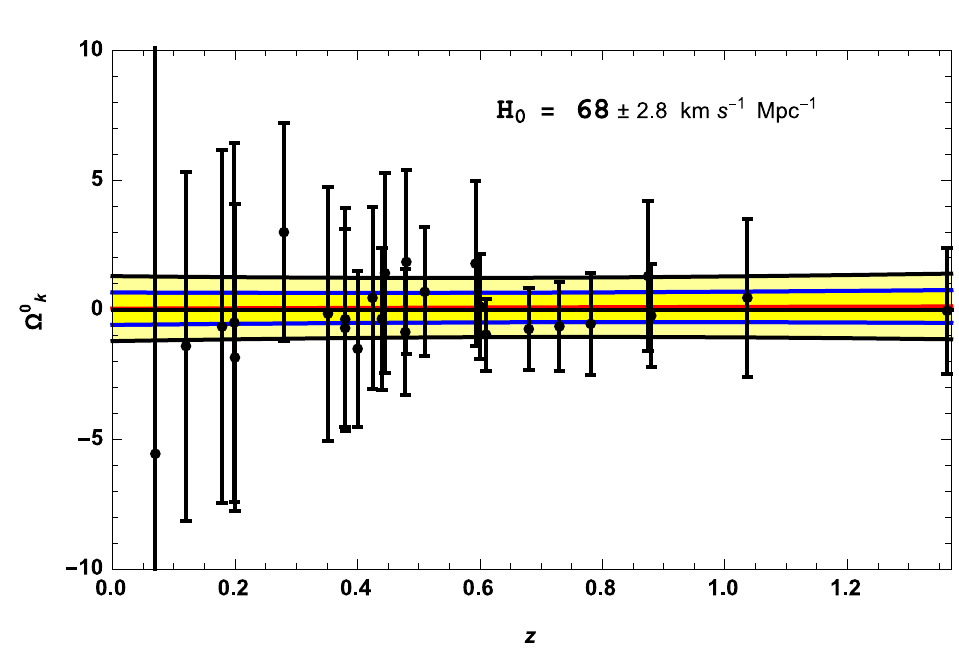}
 \includegraphics[height=5.5cm,width=7.5cm,scale=4]{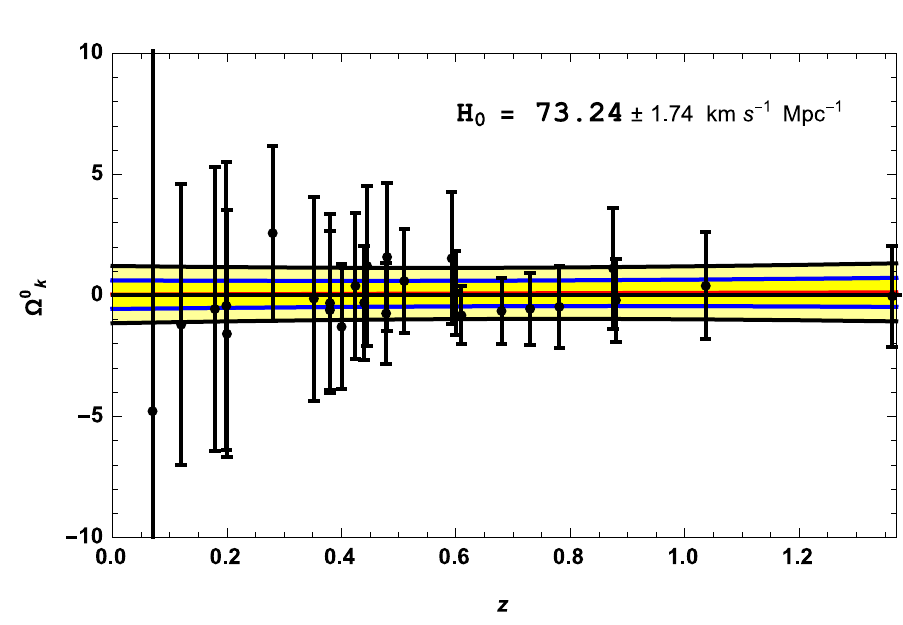}
\caption{\label{aok}\small Smooth plots of $\Omega_k^0$ vs $z$ after applying Gaussian Process for different values of Hubble constant $H_0$. The red line shows the best fit line while dark yellow and light yellow regions are the $1\sigma$ and $2\sigma$ confidence bands respectively. Black points are the $26$ reconstructed datapoints of $\Omega_k^0 $ at different redshifts. The horizontal black line represents the $\Omega_k^0= 0 $ line.}
\end{figure}

\subsection{Methodology}
In order to understand the global behavior of $\Omega_k^0$, we employ  a non-parametric method, namely Gaussian Process on the newly constructed dataset. It enables us to  obtain  a smoothened  curve with the corresponding $1\sigma$ and $2\sigma$ confidence bands. In GP method, we presume that each observation is an outcome of an independent Gaussian distribution belonging to the same population. However the outcome of observations at any two redshifts are correlated due to their nearness to each other. This correlation between two points (say $z$ \& $\hat{z}$ ) is incorporated in the technique  through a covariance function given by

\begin{center}
\begin{equation}
k(z, \hat{z})= \sigma_f^2 \exp \left[{\frac{-(|z-\hat{z}|)^2}{2l^2}}\right]
\label{3}
\end{equation}
\end{center}

In this function, the two hyperparameters, the  length-scale $l$ and  output variance $\sigma_f$, determine the length of the `wiggles' in the smoothing function and average distance of function from its mean respectively. The values of these hyperparameters is  calculated  by maximizing the corresponding marginal log-likelihood probability function of the distribution. For details see  Rasmussen et. al.(2006)\cite{rasmussen2006} \&  Seikel et.al (2012) \cite{marina2012}.

Finally, by applying the Gaussian process on the dataset of $\Omega_k^0$  obtain the best fit curves along with the $1\sigma$ and $2\sigma$ error bars.[See Fig.\ref{aok}].

\section{Method II : Constraining cosmic curvature using the mean image separation of gravitational lenses}\label{Method II}

 Gravitational lensing is an important observational tool in cosmology. This has an advantage over the other cosmological tools as it relies on relatively well understood physics. Various aspects of the gravitational lensing have been studied elaborately  which helps us to understand the geometry and the constituents of the universe. In their seminal work, TOG (1984) \cite{turner84} studied the statistical properties of gravitational lenses in order to understand the cosmological parameters and their evolution. One of the important aspects of the statistical properties is the relation between the average image separation and source redshift. In this work, we have exploited this relation to put constraints on the cosmic curvature. This analysis is based on a fact that \emph{ the mean image separation is completely independent of the source redshift for  all the FLRW based cosmological models in a flat universe i.e. $\Omega_k=0$ , if the lensing galaxy is non-evolving and modelled as a Singular Isothermal Sphere (SIS).\cite{gott89,helbig98,fukugita92,djain2004,park1997,park2014}}
In this case,  the normalized mean image separation $\frac{<\Delta \theta(z_s)>}{\Delta \theta_0}$ for the $SIS$ galaxy lenses can be defined as \cite{helbig98,park1997,park2014}

\begin{equation}\label{dt0}
\frac{<\Delta \theta(z_s)>}{\Delta \theta_0}= \dfrac{(\int\limits_0^{z_s} \frac{D_{LS}^3 D_{OL}^2 (1+z_L)^2}{D_{OS}^3 E(z_L)} dz_L )}{(\int\limits_0^{z_s} \frac{D_{LS}^2 D_{OL}^2 (1+z_L)^2}{D_{OS}^2 E(z_L)} dz_L )}
\end{equation}
where $z_L$ \& $z_S$ are the lens and source redshifts respectively.  $L$, $S$ and $O$ represent the lens, source and observer, while $D_{LS}$, $D_{OL}$ \& $D_{OS}$ indicate the corresponding angular diameter distances between the  lens , source and observer (for mathematical expressions see \cite{kayser97}). Further $E(z)= {H(z)}/{H_0}$ and  $\Delta \theta_0= \dfrac{8\pi \sigma_v^2}{c^2}$,  where $H(z)$ is the Hubble parameter and $\sigma_v$ is the velocity dispersion of the lens galaxy. \cite{keeton}.

 \begin{figure}[ht]
 \centering
 \includegraphics[height=6cm,width=8cm,scale=4]{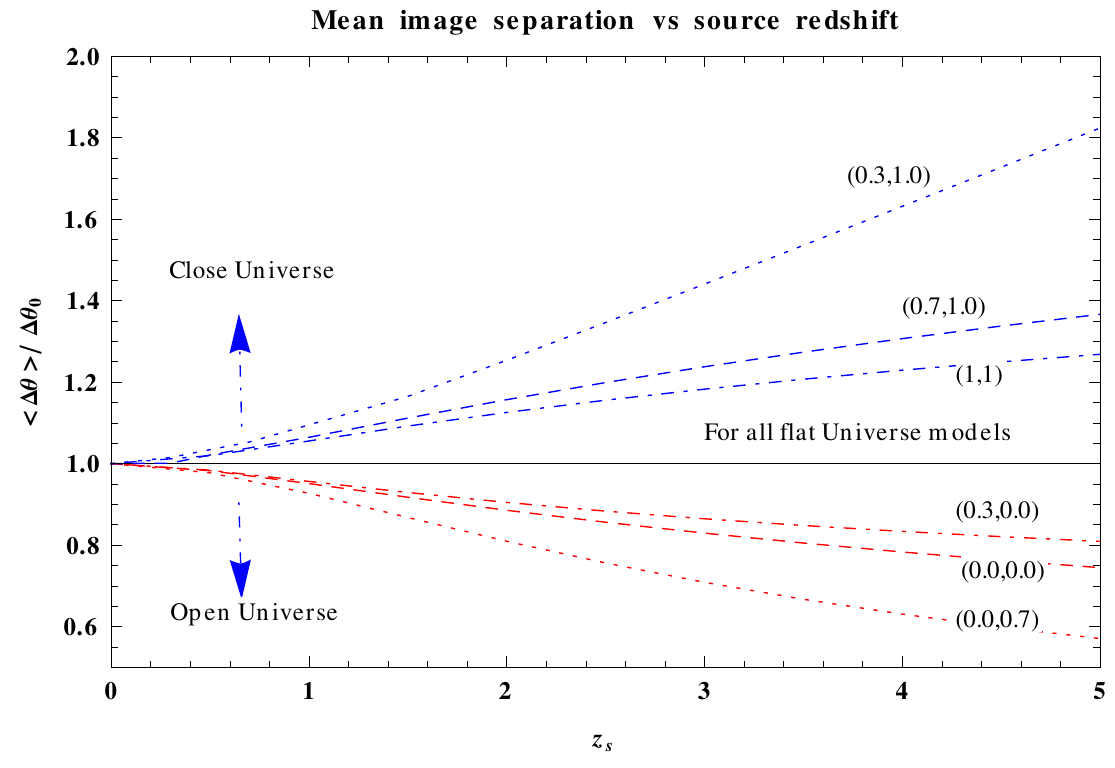}
 \caption{\label{afig} Normalized mean image separation {\small $\frac{<\Delta \theta >}{\Delta \theta_0}$ vs  source redshift $z_s$ for different cosmological models. Values in brackets represents the matter density and the  cosmological constant density i.e. $(\Omega_m, \Omega_{\Lambda}) $ configurations, while the  black solid line is for all possible flat universe configurations i.e. $\Omega_m + \Omega_{\Lambda}=1$.}} 
 \end{figure}

In Fig.\ref{afig}, we plot the normalized mean image separation $\frac{<\Delta \theta(z_s)>}{\Delta \theta_0}$ vs source redshift $z_s$. It  shows that for the SIS model of lenses the image separation integrated over the the lens redshift $z_L=0$ to $z_L=z_S$  is completely independent of the source redshift $z_s$ for a flat universe. However, the mean image separation decreases with source redshift in an open universe  and increases in a closed universe. In a closed universe the volume decreases faster with redshift than in a flat universe, and the source is more likely to be lensed by the  lensing galaxies at shorter distances, which in turn produces large image separations, and vice versa for an open universe.The aim of this work is to analyze the present available data for exploring the  possible correlation between the average image separation and source redshift. Any  correlation will  shed some light on the cosmic curvature.

\subsection{Dataset and methodology}\label{ss1s2}

We use the final statistical sample of lensed quasars from the Sloan Digital Sky Survey (SDSS) Quasar Lens Search (SQLS). This SDSS DR7 quasar catalog consists of a well-defined statistical lens sample of $26$ lenses and $36$ additional lenses identified with various techniques\cite{Inada2012}. However as we are assuming the SIS model of galaxies for lenses, this limits the number of source images to two or a perfect Einstein ring. So we select only those lens systems which have two images. Since the image separations in some lens systems are too large which can not be explained comfortably by assuming the SIS lens model \cite{lee94, park1997}. Therefore we impose a selection criterion according to which the  maximum image separation between two images should be less than $4''$ \cite{khare2001}. After applying these criteria on a dataset containing 62 lensing systems, we are finally left with $44$  galaxy lenses only. For calculating the mean image separation first we divided this dataset in  the redshift bin-size of $0.3$ and $0.5$ each and then determine  the mean value of $\Delta\theta$ in each interval (See Fig.\ref{indirectgp} upper panel). The size of the redshift bins are selected in such a way that two or more than two points should lie in each interval. Though a large bin size will increase the bias in the result, we believe  that with the availability of more data points in the near future may reduce the bias in the result.

\begin{figure}[ht]
\centering
 \includegraphics[height=6cm,width=8cm,scale=4]{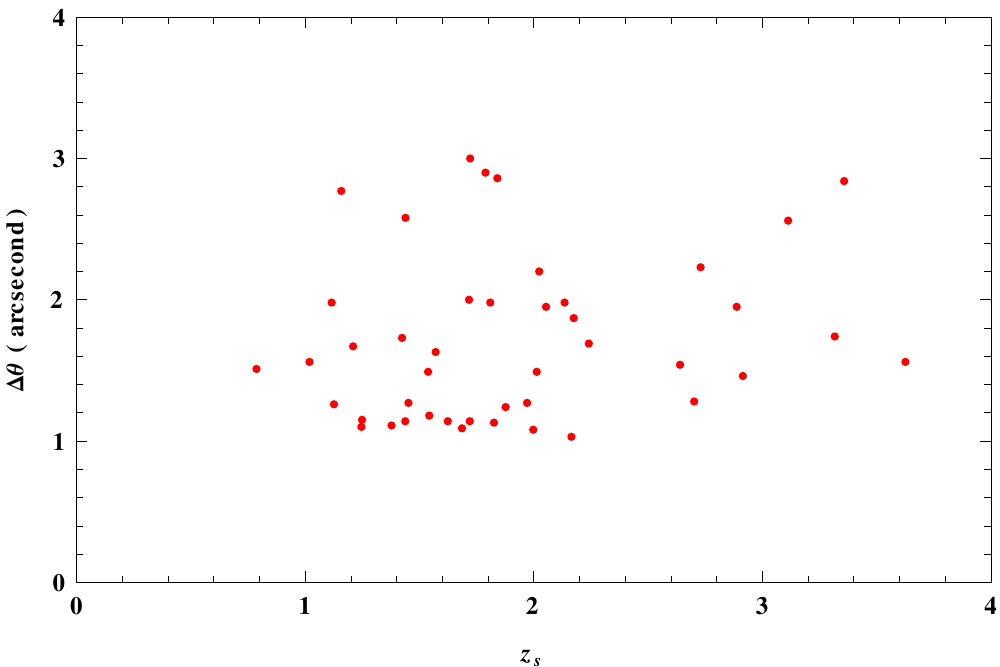}
\caption{\label{a}\small Observed data of statistical sample of lensed quasars from the Sloan Digital Sky Survey (SDSS) Quasar Lens Search (SQLS) ($44$ data points) after applying selection criteria. }
\end{figure}

In this method, our aim is to make use of the fact that the mean image separation of lensed sources  remains  constant with respect to  the source redshift for all possible flat universe models, if the lensing galaxies follow the SIS model. This enables us to differentiate between an open, closed and flat universe. In our analysis, we use the image separation of $44$ gravitational lenses and further created two datasets of  9  and 6 data points by assuming redshift bins of size $0.3$ and $0.5$ respectively. We  apply the  Gaussian process on these two datasets to study the variation of mean image separation with respect to source redshift. In the flat universe the reconstructed curves should  remain independent of source redshift as  shown in Fig.\ref{afig} and the  normalized value of the mean image separation must have a constant value equal to one. By analyzing Fig.\ref{indirectgp}, we find that the  reconstructed curves are showing a very weak inclination towards a closed  universe. However within the  $2\sigma$ confidence region,  both curves partially include the $\frac{\Delta \theta(z_s)}{\Delta \theta_0}=1$ line. Thus, we can argue  that the best fit line seems to indicate  a closed universe but within the $3\sigma$ confidence region it will accommodate the possibility of a flat universe easily.

\begin{figure}[ht]
\centering
\includegraphics[height=5.5cm,width=7.5cm,scale=4]{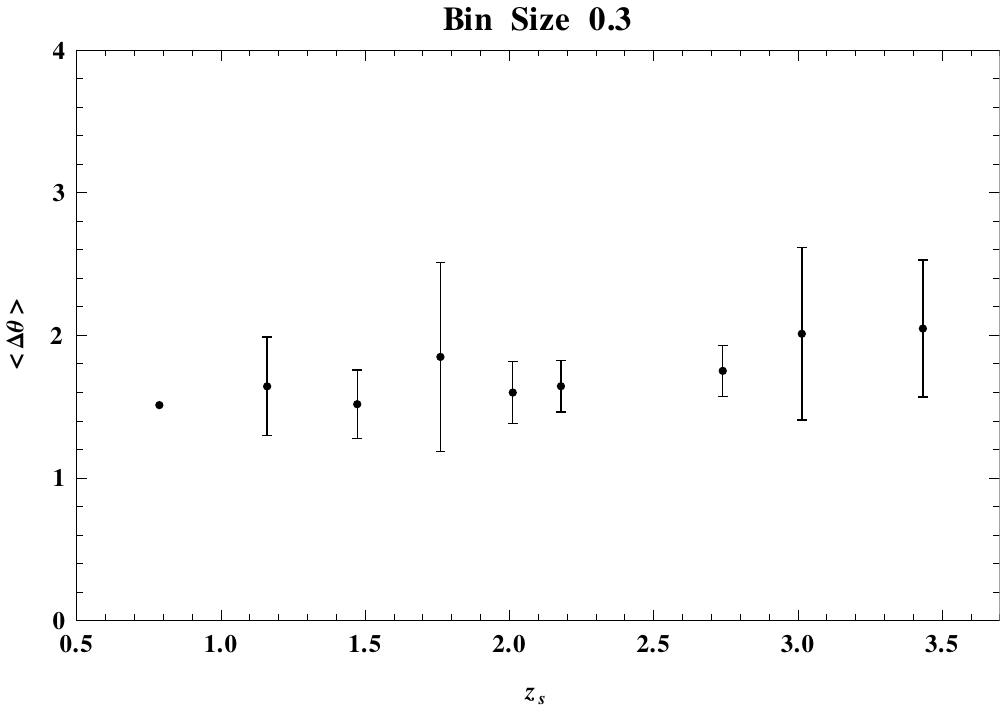}
 \includegraphics[height=5.5cm,width=7.5cm,scale=4]{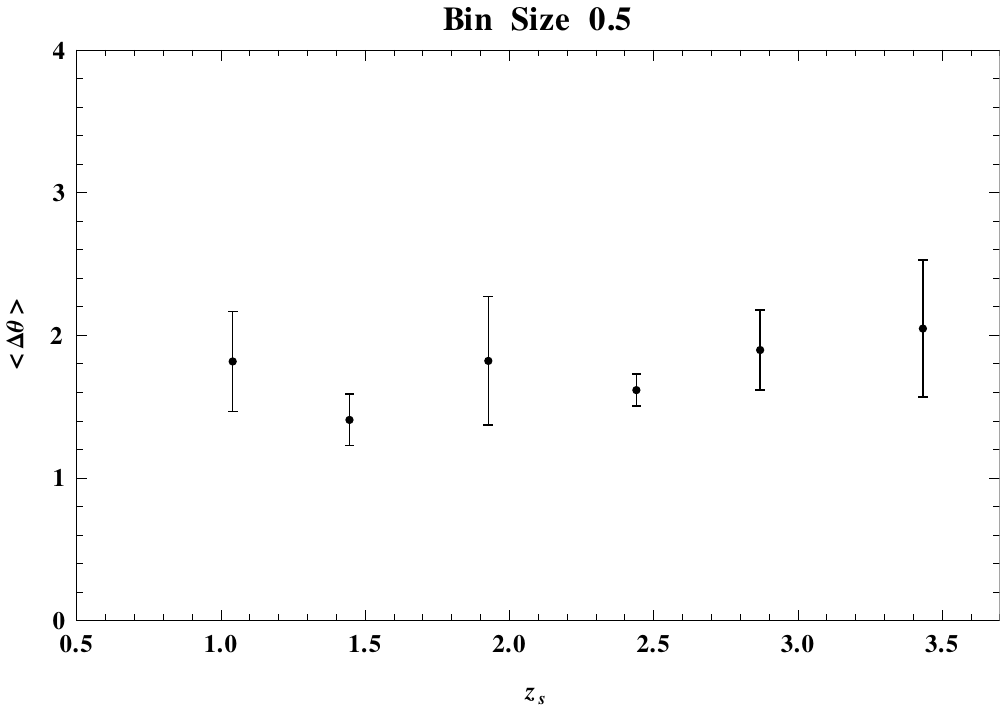}
 \includegraphics[height=5.5cm,width=7.5cm,scale=4]{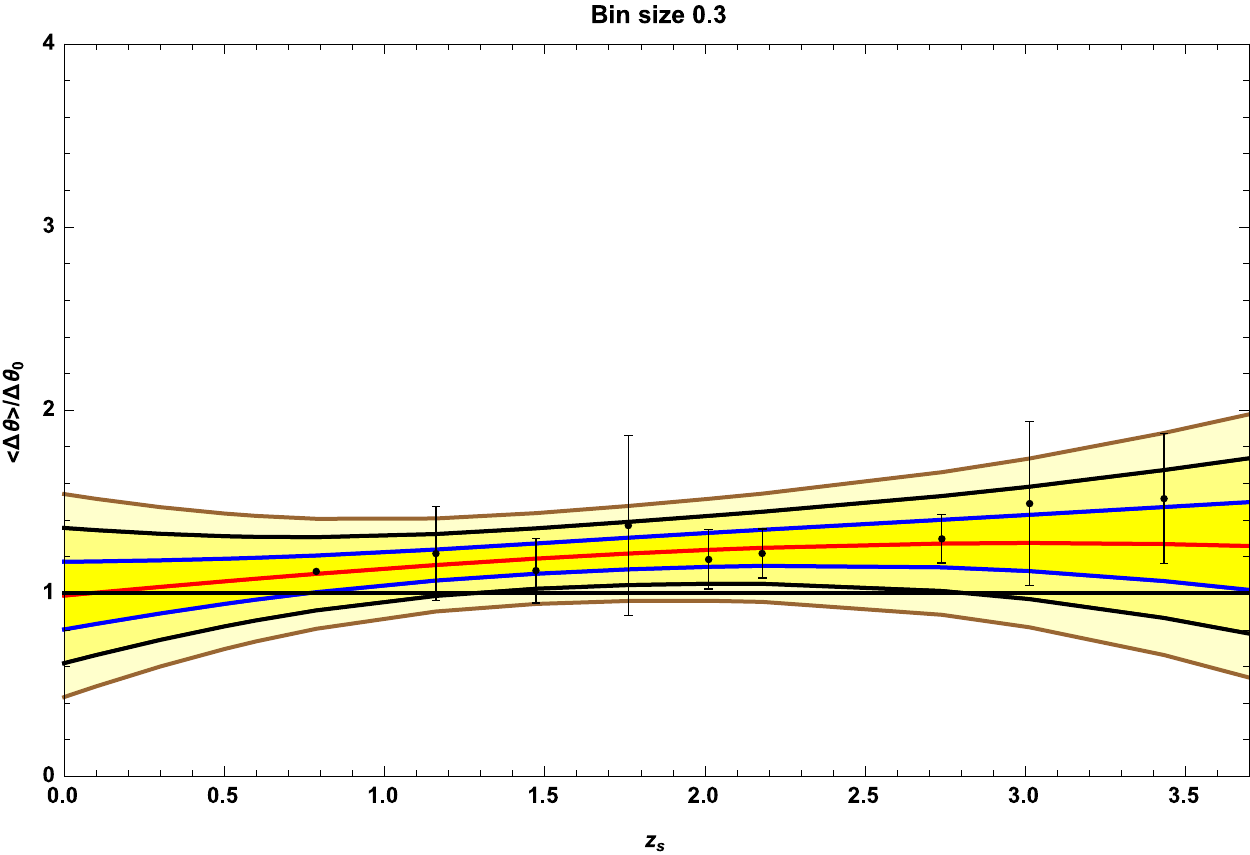}
\includegraphics[height=5.5cm,width=7.5cm,scale=4]{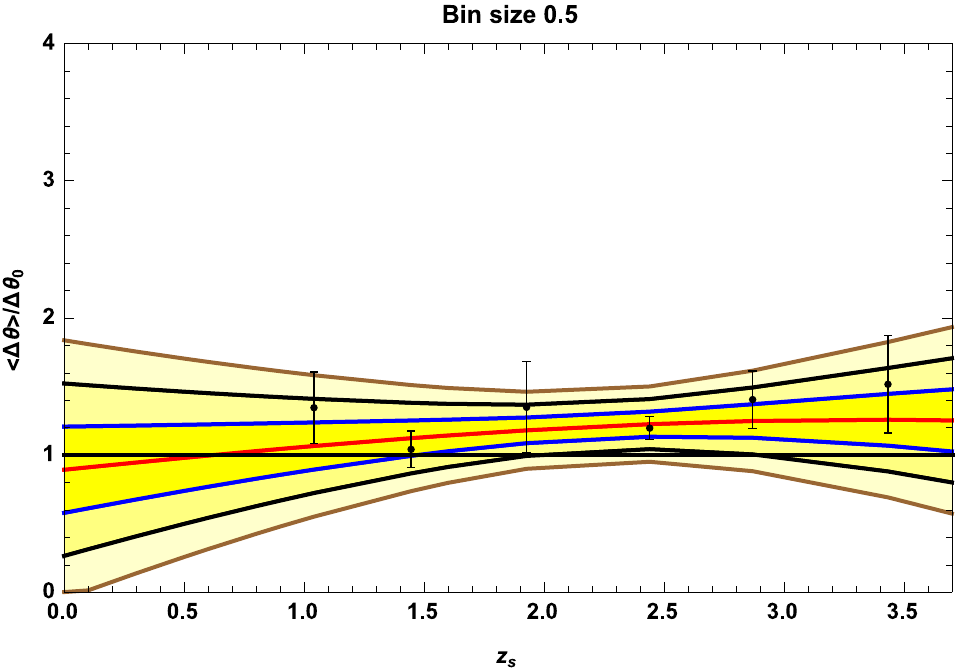}
\caption{\label{indirectgp}\small The upper two figures show the formation of different datasets using a binning size  of $0.3$ (upper left) and $0.5$ (upper right) respectively. The lower panel of figures contains the smoothened   curve of  the normalized mean image separation datasets formed using a binning size $0.3$ (left) and $0.5$ (right) respectively. The red line shows the best fit line obtained using the  Gaussian process with the $1\sigma$, $2\sigma$  and $3\sigma$ confidence regions. While  solid black line represent the theoretical variation of normalized mean image separation with source redshift $z_s$. }
\end{figure}

To further analyze this correlation between the image separation and the source redshift, we apply three statistical tests to check the correlation between parameters. Pearson product-moment correlation test, Spearman rank correlation test  and Kendall rank correlation test  are the well established tests  which are widely used to check the linearity and monotonicity or in other words, the correlation between two parameters \cite{nurec}. Out of these three tests, the Pearson product-moment correlation test gives  a parametric measure of the degree of linear dependence between two variables and is denoted by the symbol $r$. The other  two methods of Spearman's rank correlation test and Kendall rank correlation test are a nonparametric measure of the strength and direction of association that exists between two variables or check their statistical dependence  and are represented by $\rho$ and $\tau$ respectively. The coefficient corresponding to each test takes a range of values from $+1$ to $-1$.  A value of zero indicates that there is no correlation between the two variables. A value greater and smaller than zero indicates a positive  and negative correlation respectively. The value of $r$, $\rho$ and $\tau$ with the  corresponding error bars are mentioned in the Table \ref{table}. Each statistical  test indicates a weak positive correlation between image separation $\Delta \theta$ and source redshift $z_s$. These tests also shows their consistency with the variation of reconstructed curve  shown in  Fig.\ref{indirectgp}.

 \begin{table}[t]
\centering
 \begin{tabular}{|c| c||}
 \hline
  Statistical tests & value \\ [0.5ex]
 \hline
Spearman's rank coefficient ($\rho$)  	&0.22$\pm$ 0.09\\
 \hline
Pearson product-moment coefficient ($r$) 	&0.20$\pm$ 0.15\\
 \hline
Kendall rank correlation coefficient ($\tau$)	&0.13 $\pm$ 0.01\\
 [1ex]
 \hline

 \end{tabular}
 \caption{ Statistical tests to check the correlation between image separation and source redshift.}
\label{table}
\end{table}

\section{Results and Discussions}
\label{Results}

Given the crucial role played by  Cosmic curvature in the evolution of the universe, it is  important to study the cosmic curvature in a model independent manner.  Several methods have been proposed to determine the curvature parameter using different techniques. In this paper, we work  with two different approaches. Method I is a Null test of cosmic curvature implemented by using a dataset of the Hubble parameter $H(z)$ and age of galaxies. In this approach, our main focus is to check the assumption of  homogeneity  and FLRW metric of the universe. On the other hand, Method II is based on the statistical properties of gravitational lenses. It gives  us an indirect window to check the flatness of  the universe if FLRW metric of the universe holds. Hence the outcome of Method I paves the way for Method II. In totality, it gives a measure of the consistency of flat universe based on FLRW  metric. Here we will discuss the outcome of both the methods separately.

\subsection { Method I }

This technique is based on a model independent null test of cosmic curvature \cite{clarkson2007}. This test is  presented  as a test of  homogeneity of universe.  It highlights the importance of measuring the cosmic curvature in a more precise manner. The main logic behind the null test of curvature is the fact that if we have two mutually independent datasets of the Hubble parameter $H(z)$ and the transverse comoving distance simultaneously, then we can reconstruct the present value of the cosmic curvature at each redshift of measurement. In a FLRW universe $\Omega_k^0$ should not depend on the redshift and if we observe any deviation from this, then it is a direct indication of deviation from a FLRW universe or a  violation of the assumption of homogeneity of the universe  \cite{sapone,obs}. Efforts have been  made  to relate a violation of this null test of cosmic curvature to the inhomogeneous cosmology model \cite{wiltshire}. Further  Boehm et.al (2013) have presented  the violation of this null nest of curvature as a signature of  back-reaction toy model with realistic features and have tried to confront it with the Union2.1 supernova data \cite{bohem2013}.\\

The prime challenge is to check the efficiency of this null test  in a precise manner.  Earlier Union2.0 SNe Ia dataset was used for calculating the transverse comoving distance and the Hubble parameter data points are obtained from differential ages of galaxies and radial BAO measurements\cite{shafi,li2014}. This method is termed as the differential approach for constraining cosmic curvature \cite{mortsell2011}. Sapone et.al (2014) test the efficiency  of  the method with four different statistical methods \cite{sapone}, while Yu et.al \cite{yu2016} test the cosmic curvature by comparing the proper distance and transverse comoving distance. Many combinations of observational datasets have been applied on this null test but all the combinations demonstrate  their consistency with the FLRW universe and even support the flatness of universe within different confidence regions \cite{wang2016,wei2016,cai2016}.
Though many efforts have been made to measure $\Omega_k^0$  using various observational  datasets, there is always a need to use a model independent dataset. In earlier works transverse comoving distances were  frequently  obtained by using the SNe Ia dataset. This is matter of debate since calculation of distance modulus of SNeIa itself depends upon the dark energy model and also infected with systematic errors like cosmic opacity.  However in this work we obtain  the transverse comoving distance $r$ using the age of galaxies \cite{holanda2016} and combine it with the selected Hubble data points(see Table \ref{tab}) to obtain a model independent bound on  the cosmic curvature $\Omega_k^0$ at different redshifts.\\

Our approach is completely model independent  because the measurement of ages of old passively evolving galaxies relies only the detailed shape of the galaxy spectra but not on the galaxy luminosity \cite{avgo2009}. Further to study the global behavior and to constrain the $\Omega_k^0$ we apply a non parametric technique Gaussian Process and obtain a smoothened curve for it as shown in Fig.\ref{aok}. By analyzing this smoothened curve, we conclude that:

$\bullet$ The  cosmic curvature in the redshift range $0<z<1.37$ remains constant with respect to redshift for both the values of $H_0$ used in this work. This further  support a homogeneous FLRW universe. We can also infer that this null test disallows  the possibility of an inhomogeneous universe. We calculate the best fit bounds on $\Omega_k^0$ using different values of $H_0$ for a presumed flat universe [See Table \ref{hubtab}].We  repeat the same exercise (reconstruction of  $\Omega_k^0$) in open and closed universe also.  We observe that the reconstructed region completely encloses the presumed value of $\Omega_k^0$ within the  $1\sigma$ confidence level  and remains constant with respect to the redshift.  This shows consistency with  the assumption of homogeneity of the universe and also  concordance with  the FLRW metric.

\begin{table}[]
\centering
 \begin{tabular}{|c|c|}
 \hline
\small  $H_0$(Km/sec/Mpc) & $\Omega_k^0$ \\ [0.5ex]
 \hline\hline
$68\pm2.8$&$0.036\pm 0.6$2\\
$73.24\pm1.74$&$0.025\pm 0.57$\\
[1ex]
 \hline

 \end{tabular}
 \caption{\footnotesize Best fit value of $\Omega_k^0$ for different values of $H_0$   }
 \label{hubtab}
\end{table}

\subsection { Method II }

In  this method,  we  use another approach to constrain the curvature of the universe  which  is based on the  FLRW  metric. TOG ( 1984) \cite{turner84},  Gott, Park \& Lee (1989) \cite{gott89}, Fukugita et. al. (1992) \cite{fukugita92} and Lee \& Park(1994) \cite{lee94} construct the framework where statistical properties of gravitational lenses can be used to constrain the deceleration parameter and the cosmological constant.\\

Park \& Gott [PG] (1997) \cite{park1997} and Helbig \cite{helbig98} first proposed the use of statistical properties of gravitational lenses as a cosmological test for the curvature of universe. PG used a dataset of $20$ multiple-image lens systems \cite{surdej1994, keeton1996}  and observed a strong negative correlation between the image separation and the source redshift.  In order to search for the possible cause of this, Yoon \& Park (1996) \cite{yoon1996} tried to study the effect of  the evolution of lensing galaxies, clusters and curvature of the universe on the image separation of lenses.  However, Helbig \cite{helbig98} showed that in an inhomogeneous universe a negative correlation is expected regardless of the value of cosmic curvature and  argued that PG  had used an inhomogeneous sample of gravitational lenses from the literature. Khare (2001)\cite{khare2001} further extended this analysis by using a dataset of $39$ QSO lens systems  of CfA/ Arizona Space Telescope Lens Survey [CASTLES]\cite{castles} and  found no significant correlation (weak negative correlation). However recently, Xia et.al. \cite{xia2016}  use the 118 galactic-scale strong gravitational lensing (SGL) systems to constrain the cosmic curvature and find it to be close to zero.

In this work, we use the image separation data of lensed quasars in two different ways. First we divide the data into different redshift bins. After that we apply the GP smoothing technique to understand the variation of $ \Delta \theta$  w.r.t. source redshift. Second, we apply  different statistical test on the complete data of $44$ points  of image separation in order to explain the possible correlation between the two variables ( $\Delta \theta$ and $z_S$) .

The brief summary of the result is as follows:

$\bullet$ In Fig. \ref{indirectgp}, the best-fit line  which represents the  normalized mean image separation shows an  inclination towards a close universe. However, within the $3\sigma$ region it  also incorporates a flat universe. Though this inclination of  best-fit line towards a closed universe  can't  be directly  taken as the deviation from a flat universe, it does motivate us to study some of the proposed  non-flat dark energy models \cite{pavlov13,farooq13,chen16,farooqRatra16}.  Han \& Park (2014) \cite{park2014} have shown that the mean image separation in the presence of  a direct or indirect  angular selection bias or from a selection bias due to the limiting magnitude on the lens,  manifests itself as an inclination towards the  positive side with increasing source redshift. Hence, it is possible that the  deviation  of the reconstructed curve in Fig.\ref{indirectgp} is a result of the various  selection criteria  in the formation of the dataset which can give rise to  a definite bias in our result.

$\bullet$  Though the different bin sizes result in a   difference in the number of data points,  the overall trend remains the  same and  with a small bin-size we obtain comparatively tight bands. From this we can argue that in future, with the availability of more observations, the bin-sizes can be reduced  and more precise constraints can be obtained on cosmic curvature using the same method. We also believe that this test could serve as  a new window to study the cosmic curvature in a model independent way.

$\bullet$   In order to quantify the correlation,  if any,  we apply Pearson product-moment correlation, Spearman rank correlation test and Kendall rank correlation test on the  image separation versus  redshift dataset (as shown in Table \ref{table}). Contrary  to the previous work \cite{park1997, khare2001}, we observe a weak positive correlation between the variables. This result is further consistent with reconstructed curve of the normalized mean image separation vs source redshift by using GP (See Fig \ref{indirectgp}).\\

 Finally, we can conclude that Method I probes the assumptions of  a FLRW universe and homogeneity of the  universe and we observe that it confirms both the assumptions. We  also find  $\Omega_k^0= 0.025\pm0.57$ (with $H_0= 73.24\pm1.74$ Km/sec/Mpc) for a presumed flat universe. Moreover,  Method II, the indirect test  based on the image separation of gravitational lenses,  shows the consistency with a flat universe within the  $3\sigma$ confidence region (see Fig.\ref{indirectgp}). However the best fit line shows deviation from the flat universe prediction which could be due to  selection effects as explained above. Thus we can say that  both the methods used in this work jointly indicate a homogeneous but marginally closed universe. However, a flat universe can be incorporated at $3\sigma$ confidence level. In the near future with the availability of large datasets from various  surveys, we would be able to put some very tight bounds on the cosmic curvature using the same model independent and geometric tests.

\acknowledgments

 Authors are thankful to the reviewer for extremely useful comments and to Myeong- Gu Park, Phillip Helbig, Tong Jie Zhang \& Yu Hai for useful discussions.  A.R. acknowledges support under a CSIR-JRF scheme (Govt. of India, FNo. 09/045(1345/2014-EMR-I)) and also thanks IRC, University of Delhi for providing research facilities. A.M. thanks Research Council, University of Delhi, Delhi, India for providing support under R \& D scheme 2015 -16.

\end{document}